%% file: acl2023.tex
\pdfoutput=1
\documentclass[11pt]{article}
\usepackage[acl]{definition}
\usepackage{acl}

\usepackage{commands}
\usepackage[normalem]{ulem}

\newcommand{\ucla}{\includegraphics[height=0.8em]{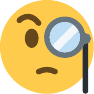}}
\newcommand{\bupt}{\includegraphics[height=0.8em]{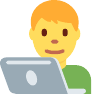}}
\newcommand{\casia}{\includegraphics[height=0.8em]{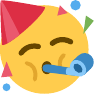}}
\newcommand{\raiseemoji}[1]{\raise0.75ex\hbox{#1}}

\renewcommand{\paragraph}[1]{\textbf{#1\,}}

\title{A Survey on Pretrained Language Models for Neural Code Intelligence}

\author{Yichen Xu\raiseemoji{\bupt} \\
EPFL\\
\texttt{yichen.xu@epfl.ch}\And
Yanqiao Zhu\raiseemoji{\ucla\,\casia} \\
UCLA \\
\texttt{yzhu@cs.ucla.edu}}

\setlist{wide,nosep}

\annotator{yichen}{red}
\annotator{yanqiao}{blue}
\newcommand{\resolved}[1]{}

\begin{document}
\maketitle

\begin{abstract}
As the complexity of modern software continues to escalate, software engineering has become an increasingly daunting and error-prone endeavor. In recent years, the field of Neural Code Intelligence (NCI) has emerged as a promising solution, leveraging the power of deep learning techniques to tackle analytical tasks on source code with the goal of improving programming efficiency and minimizing human errors within the software industry.
Pretrained language models have become a dominant force in NCI research, consistently delivering state-of-the-art results across a wide range of tasks, including code summarization, generation, and translation.
In this paper, we present a comprehensive survey of the NCI domain, including a thorough review of pretraining techniques, tasks, datasets, and model architectures.
We hope this paper will serve as a bridge between the natural language and programming language communities, offering insights for future research in this rapidly evolving field.
\end{abstract}

\input{sections/introduction.tex}

\input{sections/models.tex}

\input{sections/tasks_and_datasets.tex}
\input{sections/discussion.tex}
\input{sections/conclusion.tex}
\input{sections/responsiblenlp.tex}

\bibliography{acl2023}
\bibliographystyle{acl_natbib}

\end{document}

%% file: sections/introduction.tex
\section{Introduction}

Programming languages \cite{Pierce2002} serve as the foundation of software, enabling humans to communicate with computers and instruct them to perform computation.
The process of developing software using programming languages, known as software development, has become a thriving industry that plays a crucial role in the modern digital world. However, software development involves a range of tasks beyond programming, including testing, documentation writing, and bug fixing, which are known to be challenging and require a high level of human expertise \cite{Brooks1978}.

To ease software development, code intelligence tools have emerged as a computer-aided approach to automatically analyze source code and solve software engineering tasks.
Previously, these tools were mostly based on static analysis technologies.
For example, Microsoft Intellisense\footnote{\url{https://code.visualstudio.com/docs/editor/intellisense}} is a code intelligence tool that provides code completion suggestions and hints function signatures by statically analyzing user code and building a database of definitions, references, type signatures, and so on.
There are also tools for automatically detecting vulnerabilities in source code \cite{Ayewah2008,Engler2004}.
While these tools have been widely adopted in industry, they have limitations. One of the main limitations is that these tools are typically built for a specific programming language, requiring significant effort to migrate them to new languages. Additionally, dynamic languages like Python are difficult to analyze statically, making traditional code intelligence tools less effective for developers.

Recently, researchers have started applying language models and leveraging pretraining strategies for code intelligence tasks, such as program synthesis \cite{Chen2021,Wang2021}, documentation generation \cite{Wang2021,Alon2019b,Feng2020}, defect detection, and program repair, inspired by the success of pretrained Transformer models on modeling sequential data \cite{Krizhevsky2017,Vaswani2017}.
Motivated by the software naturalness hypothesis \cite{Hindle2016,Buratti2020}, which suggests that programming languages can be understood and generated like natural languages, researchers have treated source code as sequential data and applied sequential neural architectures, like the Transformer model \cite{Vaswani2017}, to understand and generate programs \cite{Feng2020,Guo2021}.
In the Natural Language Processing (NLP) community, it has been observed that the pretraining paradigm enables models to learn high-quality context token embeddings and significantly improves downstream performance when a large amount of unannotated data is available \cite{Krizhevsky2017,Brown2020}.
Similarly, a vast amount of code snippets for programming languages can be found on open source platforms like GitHub \cite{Chen2021}. Thus, researchers have adopted the pretraining paradigm \cite{Devlin2019} to solve various code analytical tasks \cite{Feng2020,Guo2021,Chen2021}.
Such a data-driven approach greatly saves efforts in developing code intelligent tools for supporting different tasks and adapting to new programming languages compared to static analyzers.
To date, pretrained code language models have achieved state-of-the-art performance on a wide range of tasks and demonstrated satisfactory generalization across different languages \cite{Wang2021,Chen2021,Li2022}.
Notably, Codex \cite{Chen2021} has made a significant milestone in program synthesis and has empowered programming intelligence tools like Copilot\footnote{\url{https://github.com/features/copilot}},
which have even been applied to solving linear algebra and math word problems \cite{Tang2021,Drori2021}.

Despite the success of applying pretrained language models for code intelligence tasks and the flourishing research in this field, there has been a lack of systematic reviews that categorize the growing literature.
Previous works either do not adequately consider language-based code models nor provide a comprehensive review of existing model designs, downstream tasks, and datasets
\cite{Xu2022,Wu2022,Allamanis2018survey}.
In order to model and understand the semantics of programming languages, it is necessary to discuss available datasets for both pretraining and downstream tasks and how to design appropriate neural architectures and effective training schemes.
Moreover, since source code inherently has rich structural information \cite{Xu2022,Guo2021}, how to extract the utilize the structures as prior knowledge in designing code intelligence models is a vital problem.
This requires background knowledge about the theory and technologies in the programming language community.
\citet{Wu2022ASO} survey deep learning methods for structural code understanding, with discussion on both sequence- and graph-based modeling techniques.
However, this work focuses more on the structural aspects of programs and lacks an in-depth discussion of language models and pretraining strategies.

To bridge the knowledge of both NLP and PL communities, this paper groups existing pretraining language models for code intelligence under the umbrella term \textbf{Neural Code Intelligence (NCI)} and presents a systematic review in this field.
Specifically, we review neural modeling techniques for programming languages in terms of preprocessing techniques, model architectures, and learning paradigms. Next, we discuss a variety of downstream tasks for NCI and the available datasets for training and evaluating code language models. We also explore the challenges and opportunities of applying language modeling methods for code intelligence. In addition, we maintain a curated list of NCI research, news, and tools in a GitHub repository\footnote{\url{https://github.com/Linyxus/awesome-neural-code-intelligence}}.
We hope that our work could shed light on the landscape of current research in this field, assist newcomers in understanding the recent research progress, and provide insights for future research.

%% file: sections/models.tex
\section{Pretraining Language Models for Code}

\begin{figure*}
    \centering
    \includegraphics[width=\linewidth]{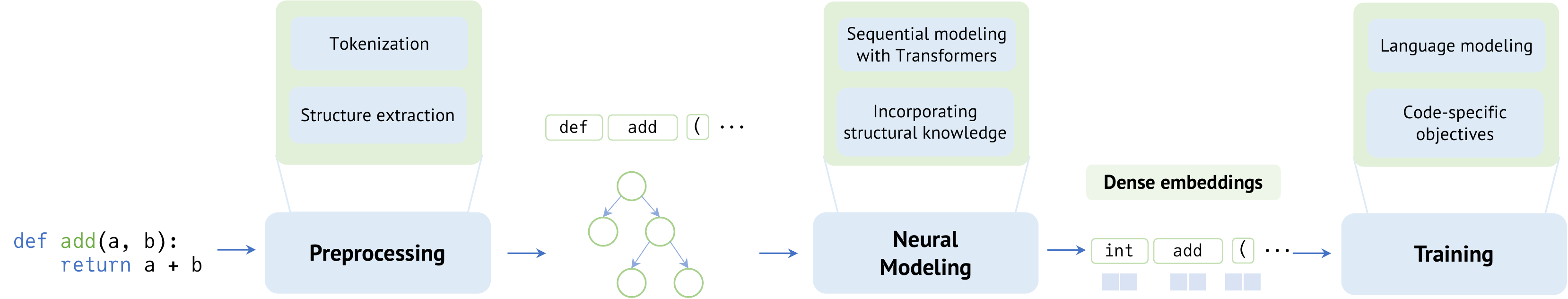}
    \caption{The pipeline of code language models: the \textbf{preprocessing} phase tokenizes code texts into sequences of code tokens and extracts code structures; \textbf{neural modeling} encodes the code tokens and structures as dense embeddings or generates output code sequences; and \textbf{training} optimizes the model with language modeling or PL-specific objectives. \resolved{Again, some words describing the whole pipeline are needed. Remove empty areas to save space. Font size should ideally be the same as \texttt{small}.}}
    \label{fig:pipeline}
\end{figure*}

In this section, we review the design and training of code language models.
To review existing literature in a systematic manner, we split the code language modeling approaches into a pipeline consisting of three stages: preprocessing, sequential modeling, and training.
The pipeline is illustrated in Figure \ref{fig:pipeline} and we will introduce each of them consecutively.

\paragraph{Preprocessing.}
For a code language model, the input will be source code snippets in the pretraining corpus or later on the downstream dataset.
It has to first preprocess the input code, including running tokenization on the code and optionally extracting prior knowledge from programming languages.

\paragraph{Sequential modeling.}
Then, it feeds the preprocessed results, mostly a sequence of tokens, into the languague model, to encode the code as dense representation, predict its property, or generate code sequences from it.

\paragraph{Pretraining and finetuning.}
Finally, we train the model with an unsupervised objective and further finetune it on downstream tasks.
In the pretraining stage, the models learns from a massive code corpus with no human annotations 
o acquire general and transferable knowledge of the structures and semantics of source code.
Then, we finetune the pretrained model for specific downstream tasks.
Apart from the aforementioned pretraining-and-finetuning strategy, other learning paradigms including zero-shot, few-shot, and multi-task learning can also be applied to train the model.

In the following sections, we review the components and different design choices of each stage.

\subsection{Preprocessing}

The preprocessing stages accepts the raw code files as input and prepares it for the next stage by running tokenization and extracting structures of the source code. 
For each sample, the output of this stage includes a sequence of tokens and optionally the structure of the source code.

We run a tokenizer on the input code files. The obtained tokens for source code are sequences of characters which group together as an elementary unit for language modeling.

Apart from tokenization, we can further optionally extract the structures from the code with the help of static analysis tools.
Programming languages are richly structured in their nature with syntax structures defined by its grammar rules and the semantic structures (e.g., flow graphs) revealed by semantic analysis tools.
The result of the structure extraction is typically represented in graphs.
For example, in a data dependency graph, nodes are variables and the edge from a node $x$ to another node $y$ means that the computation of $y$ involves $x$.

Next, we review \textbf{tokenization} strategies and \textbf{structure extraction} methods respectively.

\subsubsection{Tokenization of Source Code}

Tokenization is an indispensable preprocessing step of language models \cite{Vaswani2017}.
It splits the input text into a sequence of tokens that will be fed into the sequence model.
Previous work directly takes the tokenizers from the NLP community directly, such as BPE \cite{Sennrich2016} and SentencePiece \cite{Kudo2018}.
In addition, researchers also explore different strategies to better tokenize source code.
These strategies either improve the existing tokenizers that tailor for source code and make use of the naming conventions of programming languages and the tokenizer in their compilers to preserve the grammar and semantics of source code.

\paragraph{Fitting subword tokenizers on source code.}
The easiest way to obtain an efficient tokenizer is to fit a subword tokenizer on source code corpus.
For example, \citet{Buratti2020} propose a SentencePiece tokenizer on source code;
both CodeGPT \citet{Lu2021} and \citet{Austin2021} run the BPE algorithm \cite{Sennrich2016} on their programming-related corpus.

\paragraph{Extending tokenizers with PL-specific tokens.}
Additionally, the existing tokenizer can be improved with PL-specific tokens to encode source code efficiently.
\citet{Phan2021} add special symbols commonly seen in source code (such as \texttt{[}, \texttt{\{} and \texttt{\$}) into the SentencePiece tokenizer to better encode programming languages.
\citet{Chen2021} propose to extend a pretrained BPE tokenizer by encoding continuous whitespace of different lengths into special tokens.
According to \citet{Chen2021}, the proposed extension 
reduces the number of tokens for the same corpus and thus improves the efficiency of code language modeling.

\paragraph{Utilizing tokenizers of compilers.}
In natural languages, words are separated by spaces or punctuations, so their boundaries can be easily determined.
Unlike in natural languages that word boundaries are naturally determined by spaces and punctuations, the whitespace in programming languages is sometimes optional and the tokenizer shipped with the programming language scans tokens from source code based on a strictly-defined regular grammar \cite{compilerbook,Aho2006}.
It is thus a natural idea to tokenize the source code with PL tokenizers before running the subword tokenizer of language models \cite{Kanade2020,Roziere2020,Wang2021}.

\paragraph{Exploiting naming conventions.}
Another common preprocessing technique is to exploit naming conventions to better preserve the semantics of identifier names in the source code \cite{Svyatkovskiy2020,Peng2021b}.
Naming conventions are rules to concatenate words to form a valid program identifier.
Commonly used naming conventions are snake case (e.g. \texttt{hello\_world}) and camel case (e.g. \texttt{helloWorld}).
During the tokenization phase, we can first split identifier names based on the naming conventions before running the tokenizers so that the natural language semantics of identifiers are preserved \cite{Peng2021b}.

\subsubsection{Extracting Structures from Source Code}
\label{sec:extracting-structures-from-source-code}

In addition to improving the tokenization strategy, researchers also propose to utilize the rich structural information of source code.
Programming languages have rich syntax and semantic structures, which can be extracted using parsing tools and semantic analysis techniques.
To be specific, it is possible to extract the Abstract Syntax Tree (AST), Control-Flow Graphs (CFGs), and Data-Flow Graphs (DFGs) in the preprocessing phase and utilize these structures in code modeling \cite{Peng2021b,Shiv2019,Guo2021,Wang2021mul,Guo2022UniXcoderUC}.

To use code structures, which are often in the form of graphs or trees, in neural networks, we can \emph{flatten} them into sequences. This allows us to directly input the code structures into Transformers, which are designed to process sequential data.
One common method of flattening code structures is to obtain a sequence of nodes by traversing the graph or tree and using attention masks in the self-attention to recover the structural information, 
 which will be discussed in \cref{sec:exploiting-program-structures}.
An alternative method, proposed by \citet{Guo2022UniXcoderUC}, is to use an encoding function to map ASTs into sequences that can be directly input into the transformer without losing structural information.

\subsection{Neural Modeling for Code Tokens}

After obtaining token sequences and the optional auxiliary structures from raw input, this stage models these code token sequences with additional structural information and produces dense code representations or generates the desired programming language snippets or natural language sequences.

Existing code language models commonly include Recurrent Neural Networks (RNNs) \cite{rnn1,rnn2,Ben-Nun2018}, Long Short-Term Memory (LSTM) networks \cite{lstm,Iyer2016,Yin2017,Alon2019} and Transformers \cite{Vaswani2017}.
Among them, the majority of model extensively rely on the Transformers, the \textit{de facto} architecture for sequential modeling.
Transformers stack multiple layers of self-attention layers to model sequence data,
To exploit the rich structure of programming languages, additional components and architectures specific to language specifications are also designed.

\subsubsection{Transformers in NCI Models}

Transformers are originally designed as an encoder-decoder model. However, other variants, such as encoder-only and decoder-only models, have also been proposed in the literature. In this section, we will explore the three architectures of Transformers and their applications in the field of NCI.

\textbf{Encoder-only transformers} encode the input token sequence into dense vectors with a stack of self-attention blocks.
The resulting continuous embeddings have been widely used in code understanding models such as defect detection and code retrieval \cite{Guo2021,Feng2020,Buratti2020}.

\textbf{Encoder-decoder transformers}, such as CodeT5 \cite{Wang2021} and PLBART \cite{Ahmad2021}, have been leveraged in recent research to perform a variety of downstream tasks with prompting using multi-task training strategies similar to those used in T5 \cite{Raffel2020}.

\textbf{Decoder-only transformers}, which are extensively used in program synthesis tasks, are designed to generate high-quality sequences. Examples include Codex \cite{Chen2021} and AlphaCode \cite{Li2022}.

\subsubsection{Exploiting Program Structures}
\label{sec:exploiting-program-structures}

To incorporate the knowledge of structures in source code,
researchers propose to utilize the structural information in the self-attention module in Transformers.

\paragraph{Syntax-based self-attention.}
The self-attention module in Transformers can only model the linear position of tokens in the sequence.
While the linear modeling of sequences works perfectly for natural languages,
programming languages are structured in their nature,
and are usually parsed into Abstract Syntax Trees (ASTs) with grammar rules by compilers.
Therefore, modeling source code linearly neglects the rich syntactic structures of source code.
To model the tree structure of code syntax explicitly, it possible to modify the positional encodings in the attention module considering the syntactic structure.
For example, TPTrans \cite{Peng2021b} employs learnable positional encodings based on tree paths in the Transformer encoder.
Specifically, the tree encoding between two nodes is computed by running a Gated Recurrent Unit (GRU) network
on the path between them in ASTs.
\citet{Shiv2019} propose a stack-based absolute positional encoding scheme that can represent the node position in trees.
On the decoder side, they compute the positional encodings from the partial tree on-the-fly to ensure the alignment between outputs and the positional encodings.

\paragraph{Semantic-based self-attention.}
Source code is structured not only on the syntax level, but also on the semantic level.
Static analysis technologies can extract the semantic structures of source code, 
like the control-flow graphs and data-flow graphs.
\citet{Guo2021} propose to incorporate data-flow structures into the attention module.
They first input the data flow graph as sequence together with the source code and
then mask the attention based on the edges in the data flow graph and the correspondence between identifiers in the source code and the nodes in the graph.

\subsection{Training}

\subsubsection{Language Model Pretraining}

Most code language models follow the pretraining paradigm.
The pretraining objective is a key ingredient here.
We first review the objectives that are based on language modeling in the NLP community,
then discuss the PL-specific objectives utilizing the prior knowledge of PL.

\paragraph{Language modeling objectives.} 
Here, we briefly discuss the pretraining objectives that are widely used in natural language models and explain how to adapt these objectives for source code.

\begin{itemize}
\item \textbf{Masked Language Modeling (MLM).}
Inspired by the cloze task \cite{Taylor1953},
the MLM objective is used to pretrain BERT \cite{Devlin2019} in NLP research,
where we first mask out a portion of tokens in the input sequence and then ask the model to predict them.
A variety of encoder-only \resolved{encoder-only not introduced before!} code language models utilize MLM objective to pretrain the model \cite{Buratti2020,Kanade2020,Guo2021,Feng2020}.

\item \textbf{Next Sentence Prediction (NSP).}
The NSP objective also originates in BERT \cite{Devlin2019}.
In each sample, we randomly select two sentences $\bm s$ and $\bm t$,
where $\bm t$ can either be the real next-sentence of $\bm s$ or sampled from other places.
The sample is organized as $\tCls$, $\tokSeq{s}$, $\tSep$, $\tokSeq{t}$, $\tSep$ to be fed into the model.
The final embedding of \tCls{} token is regarded as the sequence embedding and is used to predict whether $\bm t$ is a real next-sentence.
When applying NSP on programming languages \cite{Kanade2020,Liu2020}, \citet{Kanade2020} proposes to use the logical lines in source code instead of the physical lines to better model the syntax of programming languages.

\item \textbf{Masked Span Prediction (MSP).}
The MSP obejctive is widely applied to train language models with encoder-decoder architectures \cite{Raffel2020}, which randomly masks spans of tokens in the input sequence and then predicts these masked spans combined with some sentinel tokens on the decoder side.
Code language models with a encoder-decoder architecture mostly train the model with NSP \cite{Wang2021}.

\item \textbf{Unidirectional Language Modeling (LM).}
Many decoder-only language transformers \cite{Brown2020} leverage the LM objective,
where the model is to predict future token based on the previous sequence.
Specifically, when using the objective to train Transformers, we use casual attention masking, where each token in the sequence can only attend to the tokens before it.
Then, on the output layer, we employ a classification head on each output token embeddings to predict the next token the input sequence and use cross-entropy loss to optimize the model.
In NCI, code synthesis models based on GPT employ the LM objective \cite{Chen2021,Austin2021}.

\item \textbf{Denoising AutoEncoding (DAE).} 
DAE is applied to train seq2seq language models \cite{Lample2018,Roziere2020}.
It corrupts the input sequence by randomly masking, removing, and shuffling the tokens and enforces the model to recover the original sequence from the corrupted one.
Note that the DAE loss can be regarded as an extension to the MLM loss, since it not only masks the input tokens but also permutes and drops them.
DAE is employed to train code translation models like TransCoder \cite{Roziere2020} and is shown to enhance the capability of the model to decode the internal representation of code \cite{Roziere2020}.

\item \textbf{Back translation.}
Back translation is a pretraining objective designed for cross-lingual language models \cite{Lample2018}.
\citet{Roziere2020,Roziere2021} employ this objective to train programming language models that is capable of translating between different programming languages.
Specifically, for an input code sequence of language $A$, we first ask the model to translate it to the equivalent code sequence of language $B$.
Then, we enforce the model to translate the sequence in $B$ back to the original sequence in $A$.

\end{itemize}

\paragraph{Code-specific objectives.}
Apart from objectives originated from language modeling, researchers also explore pretraining objectives specific for programming languages.
These objectives make use of the prior knowledge of programming language and software engineering domains and are proven to achieve performance improvements in many NCI tasks \cite{Roziere2021}.
    
\begin{itemize}
    
\item \textbf{Replaced Token Detection (RTD).}
Inspired by \citet{Clark2020}, the RTD objective randomly replaces tokens in the input sequence with several tokens by a generator and trains a binary classifier to predict whether a token has been replaced.
It has been employed to train BERT-like code language models \cite{Feng2020}.

\item \textbf{Identifier DeOBFuscation (DOBF).}
Inspired by the notion of identifier deobfuscation in software engineering, the DOBF objective \cite{Roziere2021} first masks the function and variable names with placeholder tokens and then train the model to recover the original names from the placeholders.
DOBF has been shown to marginally improve model performance in code translation tasks \cite{Roziere2021}. %

\item \textbf{Data-flow-based objectives.}
GraphCodeBERT \cite{Guo2021} incorporates data flow information of the input source code.
Data flow is a graph of variable dependencies in the source code, 
where each node represents a variable and each edge means that the computation of one variable depends on another one.
\citet{Guo2021} feed the data flow of source code into model input and design two data-flow-based pretraining objectives in their work,
namely \textbf{Edge Prediction} and \textbf{Node Alignment}.
Edge Prediction requires the model to predict masked edges in the data flow graph,
while Node Alignment asks the model to predict the alignment between identifier tokens in the source code and the nodes in the data flow graph.

\item \textbf{Contrastive Learning (CL).}
Contrastive learning trains the model by maximizing the similarity between positive samples. Specifically, it first generates various samples from each input data sample with data augmentations.
For each input sample, its augmented samples are regarded as positive samples, while other samples are treated as negative ones.
Then, CL optimizes the model by enforcing it to discriminate between positive and negative pairs.
Recently, ContraCode \cite{Jain2021} adapts the idea of CL to train source code language models.
ContraCode employs semantic-preserving code transformations as data augmentations and maximizes the agreement between embeddings of augmented programs with identical functionalities.
UniXCoder \cite{Guo2022UniXcoderUC} proposes a multi-modal contrastive learning objective that learns code semantic embeddings.

\end{itemize}

\subsubsection{Additional Learning Schemes}

\paragraph{Zero- and few-shot learning.}
It is found that without finetuning on downstream tasks, large pretrained models are already capable of completing various tasks such as sentiment analysis, passage summarization, and question answering \cite{Brown2020}.
This is similar for code language models.
\citet{Chen2021} propose Codex, a 12B code GPT pretrained on hundreds of gigabytes of Python source code.
It can achieve high accuracy on the program synthesis task without finetuning.
They also find that Codex largely outperforms finetuned counterparts in few-shot settings.
Surprisingly, Codex also has strong performance in solving linear algebra problems and math-word problems \cite{Drori2021,Tang2021} in zero-shot settings, even if it was not optimized for any of these problems.
Moreover, researchers find that pretrained code models show promising performance for other downstream tasks such as neural execution, type inference, variable naming, and docstring generation.

\paragraph{Multi-task learning.}
Instead of training different models for different tasks, the multi-task learning paradigm trains a unified model on multiple downstream tasks \cite{Raffel2020}.
An example is CodeT5, which is a T5-based code language model \cite{Wang2021}.
In the finetuning phase, it formalizes downstream tasks as several seq2seq problems and constructs prompts to hint the task type in the input.
Another example MulCode \cite{Wang2021mul} has dedicated input and output layers for each downstream tasks and has a unified representation layer for modeling the source code.
The whole model is optimized jointly on multiple downstream tasks.

%% file: sections/tasks_and_datasets.tex
\section{Tasks and Datasets}

In this section, we first review downstream NCI tasks and datasets in two categories: understanding and generation tasks.
Then, we review code corpus for pretraining code language models.

\subsection{Downstream Tasks}

Downstream tasks can be summarized from four dimensions: \textbf{task types}, \textbf{inputs and outputs}, \textbf{evaluation}, and \textbf{available datasets}.
We first present a taxonomy of downstream tasks, explaining the meaning of each dimension, followed by detailed introduction to each task.
Finally, we review the datasets available for each downstream task and disucss widely-used pretraining corpus for code language models.
The downstream tasks are summarized in Table \ref{tab:task-taxonomy}.

\begin{table*}
	\centering
	\resizebox{\textwidth}{!}{%
	\begin{tabular}{ccccll}
	\toprule
	Name & Type & Input & Output & Evaluation & Datasets \\
	\midrule
	Clone detection & U & PL & Binary label & Acc & \citet{Svajlenko2014} \\
	Defect detection & U & PL & Binary label & Acc & \citet{Zhou2019}, \citet{d2a}  \\
	Code retrieval & U & NL, PL & Score & MRR, \sAtK{} & \citet{Husain2019} \\
	Code summarization & G & PL & NL & BLEU, ROUGE & \citet{Lu2021}, \citet{LeClair2019} \\
	Program synthesis & G & NL, PL & PL & BLEU, CodeBLEU, functional correctness & \citet{Iyer2018}, \citet{Chen2021}, \citet{Li2022} \\
	Code refinement & G & PL & PL & Acc & \citet{Tufano2019} \\
	Code translation & G & PL & PL & Acc, BLEU, CodeBLEU & \citet{Lu2021} \\
	Code completion & G & PL & PL & Acc, Edit Sim & \citet{Lu2021} \\
	\bottomrule
	\end{tabular}%
	}
	\caption{Representative Neural Code Intelligence (NCI) tasks. Each task is characterized by its type (Understanding (U) or Generation (G)), input and output (Programming Language (PL) and Natural Language (NL)), evaluation protocols, and available datasets.
	}
	\label{tab:task-taxonomy}
\end{table*}

\begin{itemize}
\item \textbf{Task types.}
Each downstream tasks can be categorized into two classes: \textbf{Understanding tasks (U)} and \textbf{Generation tasks (G)}.
An understanding task usually takes natural language and programming language sequences as input and requires the model to predict the corresponding label.
To complete such tasks, a model only needs to \emph{understand} the meanings of natural language and source code.%
By contrast, the generation tasks will require the model to generate programming language or natural language texts from the input.
Apart from the class, we also specify the input and output of each task.
The inputs and outputs are usually \textbf{Natural Languages (NL)}, \textbf{Programming Languages (PL)}, and their labels.

\item \textbf{Evaluation protocols.}
Understanding tasks are typically formulated as a classification or retrieval problem, so the traditional metrics can be applied as evaluation metrics.
For generative tasks, we measure the performance in terms of the textual and structural similarity with ROUGE, BLEU, and CodeBLEU \cite{bleu,rouge,codebleu}, or test the functional correctness of the generated programs \cite{Chen2021}.

\end{itemize}

\subsubsection{Understanding Tasks}
\label{sec:understanding-tasks}

\paragraph{Clone detection \cite{Svajlenko2014}}
involves identifying plagiarism between two code snippets by measuring their semantic similarity. The performance of clone detection algorithms can be evaluated using metrics such as accuracy.
BigCloneBench \cite{Svajlenko2014} and CodeNet \cite{codenet} are two representative benchmarks for evaluating the performance of code clone detection algorithms.

\paragraph{Defect detection \cite{Zhou2019}}
aims to predict the vulnerability of given source code, or whether it contains bugs and defects that may result in runtime errors or attacks.
The input for this task is typically code snippets and the output is a binary label indicating vulnerability. 
As a binary classification problem, its performance can be evaluated in accuracy.
Devign \cite{Zhou2019} a defect detection dataset created from four large-scale open-source C projects, comprising 48,687 samples with diverse vulnerabilities.
D2A \cite{d2a} is another defect detection dataset created by analyzing code before and after commits, providing 1,295,623 samples.

\paragraph{Code retrieval \cite{Husain2019,Ling2021}}
which is also known as semantic code search, aims to retrieve code snippets relevant to a given natural language query.
The task can be naturally formulated as a matching problem: the model will be given a pair of NL and PL sequences representing the query and the candidate code snippets. Then, the model should output a relevance score between the query and each of the candidate snippets.
Ranking metrics like MRR and \sAtK{} can be used to evaluate model outputs \cite{Ling2021}. 
CodeSearchNet Challenge \cite{Husain2019} is a code retrieval dataset built from open source repositories of six popular programming languages.

\subsubsection{Generation Tasks}
\label{sec:generation-tasks}

\paragraph{Code summarization \cite{Lu2021,LeClair2019}}
generates explanatory natural language documentation from the given source code snippets.
The output natural language description can be evaluated with BLEU and ROUGE scores.
The CodeSearchNet corpus \cite{Husain2019} (which comes with the CodeSearchNet Challenge dataset) can be used as the code summarization dataset \cite{Lu2021}.
FunCom \cite{LeClair2019} is a code summarization benchmark of over two million Java method paired with one-line natural language description.

\paragraph{Program synthesis \cite{Chen2021,Li2022}}
or known as code generation, requires the model to generate programs from text specifications.
The input of this task is usually a program specification, which can be natural language descriptions, pseudo-code, or example input-output test cases.
These specifications can be represented in the form of natural language, programming language or the combination of the two.
In addition to textual similarity scores like BLEU and ROUGE, generated programs can also be evaluated with CodeBLEU and functional correctness, which will take the syntactic and semantic aspects of PL into consideration.

There are a variety of available datasets for this task.
CONCODE \cite{Iyer2018} is a Java code generation dataset scraped from GitHub where each sample asks the model to generate source code snippet from natural language query in programming contexts.
HumanEval \cite{Chen2021} is a program synthesis benchmark that consists of hundreds of hand-written coding challenges.
CodeContests \cite{Li2022} and APPS \cite{Hendrycks2021} are program synthesis datasets extracted from online coding platforms like Codeforces\resolved{\url{https://codeforces.com/}}, CodeChef\resolved{\url{https://www.codechef.com/}}, and Codewars\resolved{\url{https://www.codewars.com/}}.

\paragraph{Code refinement \cite{Tufano2019,Drain2021}}
aims to fix bugs in the given code snippet and output a bug-free one.
This task can be viewed as an extension of defect detection, in the sense that code refinement has to not only find the defect but also fix it.
The output is usually evaluated with the exact matching accuracy (i.e. the predicted fix matches with the ground truth exactly), since code refinement is a correctness-critical task.
\citet{Tufano2019} propose a code refinement dataset mined from bug-fixing commits in thousands of Java projects hosted on GitHub.

\paragraph{Code translation \cite{Lu2021,Roziere2020}} translates code snippets in one language to another, which is useful for migrating code bases to a new language.
Similarly, the translated code can be evaluated with accuracy, BLEU, and CodeBLEU.
CodeTrans \cite{Lu2021} is a Java-C\# code translation extracted from parallel functions in several open-source projects.

\paragraph{Code completion} aims to complete partial code snippets.
\citet{Lu2021} propose a code completion benchmark dataset built from PY150 \cite{Raychev2016} and GitHub Java corpus \cite{Allamanis2013}.
The benchmark contains two subtasks: token- and line-level completion, which require the model to predict the next single token or next line of tokens respectively.
The token level completion is evaluated with accuracy; 
the line-level prediction can be evaluated with edit similarity as in CodeXGLUE \cite{Lu2021}.

\subsection{Pretraining Corpus}

Now, we introduce several representative code pretraining corpus.
We summarize commonly-used pretraining corpus in \cref{tab:dataset}.

\begin{table}
	\centering
	\small
    \resizebox{\linewidth}{!}{
	\begin{tabular}{cccc}
	\toprule
	Name & Language & Size & Content \\
	\midrule
	GitHub Java & Java & 3 GB & {Java code files} \\
	PY150 & Python & $\approx$ 350 MB & Abstract syntax trees \\
	CodeSearchNet & Multiple & --- & Code-description pairs \\
	The Pile & Multiple & 630.64 GB & Code files\\
	CodeParrot & Python & 180 GB & Python code files \\
	The Stack & Multiple & 3.1 TB & Code files \\
	\bottomrule
	\end{tabular}
    }
	\caption{Statistics of representative pretraining corpus.}
    \label{tab:dataset}
\end{table}

\paragraph{GitHub Java corpus \cite{Allamanis2013}} is constructed from 14,807 Java projects on GitHub.
It contains 350 million lines of code and roughly 1.5B of code tokens, summing up to more than 3 gigabytes.

\paragraph{PY150 \cite{Kanade2020}} contains parsed abstract syntax trees of Python programs scraped from GitHub.
The train split consists of 100,000 files and the evaluation split has 50,000 files.

\paragraph{CodeSearchNet corpus \cite{Husain2019}} is built from GitHub projects in multiple mainstream programming languages.
It contains 2 millions of functions parsed from the source code, each paired with its natural language documentation string.

\paragraph{The Pile \cite{pile}} is an open-source English corpus for pretraining general purpose language models.
It contains 630.64 GB of source code text in multiple programming languages.

\paragraph{CodeParrot \cite{transformerbook}} is a Python code corpus built from public GitHub repositories available on Google's BigQuery.
After preprocessing, it results in a dataset containing 180 gigabytes of Python code from over 20 million source files.

\paragraph{The Stack \cite{thestack}} is a 3.1TB source code dataset consisting of permissively licensed code in 30 programming languages.
The dataset is collected from GitHub, after license filtration and near-deduplication.

%% file: sections/discussion.tex
\section{Challenges and Opportunities}

Despite the blossom development of this field, there are still many outstanding challenges and open problems.
Next, we discuss possible opportunities for follow-up research.

\paragraph{Exploiting prior knowledge of source code.}
Programming languages are well-structured by its nature:
we can parse source code into abstract syntax trees with the grammar rules;
also, there are a variety of static analyzers to extract the semantic structures of programs, ranging from type checkers to flow analyzers.
Although prior work makes a few attempts in exploiting these knowledge \cite{Peng2021,Shiv2019,Guo2021,Wang2021mul}, there is still a lack of principled way to inject structure bias in language models designed for sequential data.
Additionally, programs can be executed, revealing the runtime semantics of programs, which is the essential property of source code for various tasks, in particular program synthesis \cite{Chen2021,Li2022}.
Most previous work \cite{Chen2021,Simmons-Edler2018} focuses on machine-level instructions and cannot generate general-purpose programs with high-level programming languages.
Utilizing such rich prior knowledge for NCI models, especially the runtime semantics is a promising research direction in the future.

\paragraph{Linking project- and library-level knowledge.}
The input and output of existing code language models are mostly functions, classes, and standalone code snippets.
By contrast, real-world source code projects have a hierarchy of source code modules that have to be considered simultaneously.
Existing methods are limited to understand and generate source code in a single file and lacks the capability of modelling interconnected code modules, which is essential for NCI methods to scale up to modeling real-world code projects.

\paragraph{Broadening the horizons of NCI models.}
Researchers have discovered that code language models have applications beyond understanding and generating source code. For example, program synthesis models have been used to solve linear algebra problems, math word problems, and perform automated proof search \cite{Tang2021,Drori2021,Polu2022}.
Specifically, \citet{Polu2022} find that code language models can generate mathematical proofs, with their GPT-f model capable of proving theorems at the International Mathematical Olympiad level \cite{Polu2022}.
Codex has also been utilized in computer science education, such as solving code problems, explaining code snippets, and generating course materials \cite{Finnie-Ansley2022,MacNeil2022AutomaticallyGC}.
ProgPrompt \cite{Singh2022ProgPromptGS} demonstrates that code language models can generate task plans for robots.
In addition, it is likely that code language models could be applied to other tasks such as accelerating SAT solvers, parsing NL into SQL queries, and solving a broader range of math problems. Applying code models in these domains is an emerging but exciting area of research.

%% file: sections/conclusion.tex
\section{Conclusion}

In this survey, we develop a systematic review of the tasks, datasets, and methodology of language-modeling-based Neural Code Intelligence (NCI) models.
Our review begins with an overview of NCI tasks and datasets, followed by a detailed analysis of existing language-modeling methods.
We also delve into the challenges and opportunities presented by this rapidly evolving field. We hope that our work will provide both novice and experienced researchers with a clear understanding of the current state of NCI research and offer insights into future trends and directions.

%% file: sections/responsiblenlp.tex
\section{Limitations and Social Impacts}

\subsection{Limitations}
As a survey of the neural code intelligence field, one main limitation of this paper is that it focuses primarily on sequential-modeling-based models, while structural models are only briefly mentioned. While this is a deliberate decision in light of the scope of this survey, it is worth noting that structural models, which heavily rely on prior knowledge of code semantic structures, form a significant portion of NCI research and deserve further discussion.
Another limitation is that this survey does not address ethical and liability issues related to code language models. For example, large code language models such as Codex \cite{Chen2021} and AlphaCode \cite{Li2022} are often trained on large code corpora collected from open source communities. There are concerns that using open source code to train language models may violate open source licenses \cite{fsf-copilot}.

\subsection{Social Impacts}
Our work has both positive and potentially negative impacts on society. On the positive side, our survey illustrates the current state and envisions future directions of applying language modeling to Neural Code Intelligence (NCI), which can help guide subsequent research in this field. By advancing the research in NCI, we can improve the reliability and efficiency of the software engineering industry and enable humans to build software more efficiently.
However, there is a potential negative impact as well. This survey and much of the NCI research community have not adequately addressed ethical and open source license issues. Code language models are trained on source code produced by humans, and it is thus important to respect the rights of the programmers who wrote the code.